\documentclass[12pt]{article}
\usepackage[T1]{fontenc}

\makeatletter

\providecommand{\LyX}{L\kern-.1667em\lower.25em\hbox{Y}\kern-.125emX\@}

\usepackage{amssymb}
\usepackage{amsmath}  
\usepackage{amsbsy}
\numberwithin{equation}{section}

\newcommand{\dis}{\displaystyle}

\renewcommand{\hat}{\widehat}
\renewcommand{\tilde}{\widetilde}
\newcommand{\bs}[1]{\boldsymbol{#1}}
\newcommand{\bc}{\bs{c}}
\newcommand{\bx}{\bs{x}}
\newcommand{\bff}{\bs{f}}
\newcommand{\bg}{\bs{g}}
\newcommand{\bh}{\bs{h}}
\newcommand{\bt}{\bs{t}}
\newcommand{\bn}{\bs{n}}
\newcommand{\lr}[1]{\langle{#1}\rangle}
\newcommand{\bcp}{{\bc '}}
\baselineskip=\normalbaselineskip

\makeatother

\begin{document}

{\par\raggedleft PUPT-1933\par}

{\par\centering \textbf{\LARGE Loop dynamics and AdS/CFT correspondence}{\LARGE \vspace{1cm}}\LARGE \par}

{\par\centering {\large A. M. Polyakov\( ^{1} \) and V. S. Rychkov\( ^{2} \)\vspace{1cm}}\large \par}

{\par\centering \( ^{1} \) Joseph Henry Laboratories, Princeton University,
Princeton, NJ 08544\par}

{\par\centering polyakov@viper.princeton.edu\par}

{\par\centering \( ^{2} \) Department of Mathematics, Princeton University,
Princeton, NJ 08544\par}

{\par\centering rytchkov@math.princeton.edu\par}

\section*{\centerline{Abstract}}

We consider the strong coupling limit of conformal gauge theories in 4 dimensions.
The action of the loop operator on the minimal area in the AdS space is analyzed,
and the Schwinger-Dyson equations of gauge theory are checked. The general approach
to the loop dynamics developed here goes beyond the special case of conformal
theories.

\noindent \vspace{1cm}

\noindent May 2000

\pagebreak

\section{Introduction}

In spite of the great recent progress in understanding gauge fields\,--\,strings connection
and its special case---AdS/CFT correspondence, there is still no true derivation
of these phenomena based on the first principles. In the paper \cite{1} we
made a step in this direction by demonstrating that the loop equations of gauge
theory can be verified on the string side in WKB approximation and for the special
contours (wavy lines). 

In this note we will give a general treatment of the problem valid for any loops.
Our new approach turns out to be simpler. It reveals some amazing features,
connecting the loop Laplacian and the minimal area functional in the AdS space.
As a result, we will be able to check the loop equations in WKB approximation
for arbitrary non-selfintersecting loops. There is a direct path from here to
the full quantum theory, but it will be left for the future, apart from several
comments.

Let us first formulate our main results. Consider the Wilson loop
\begin{equation}
\label{1}
W[C]=\frac{1}{N}\left\langle \textrm{Tr}P\exp \oint _{C}A_{\mu }dx^{\mu }\right\rangle \; ,
\end{equation}
 where we average the ordered exponential over the Yang-Mills fields. In this
pure gauge theory (\ref{1}) satisfies the loop equation
\begin{equation}
\label{2}
\hat{L}(s)W[C]=W*W\: ,
\end{equation}
 where the RHS is zero for non-selfintersecting loops and the loop Laplacian
\( \hat{L} \) is defined by 
\begin{equation}
\label{3}
\frac{\delta ^{2}W}{\delta c_{\mu }(s)\delta c_{\mu }(s')}=\left( \hat{L}(s)W\right) \cdot \delta (s-s')+\text {non-local\, terms}\: .
\end{equation}
 The decomposition (\ref{3}) is not always possible---its existence is characteristic
of the zigzag-invariant functionals \cite{1}. 

When there are other fields in the theory (like in \( \mathcal{N}=4 \) SYM)
one can define many different loop functionals, like the one introduced in \cite{2},
\cite{3}
\begin{equation}
\label{4}
\widetilde{{W}}[c(s),y(s)]=\frac{1}{N}\left\langle \textrm{Tr}P\exp \oint _{C}\left( A_{\mu }dx_{\mu }+\Phi _{a}dy_{a}\right) \right\rangle \; .
\end{equation}
 In principle we can integrate around the loop any operator in the adjoint representation
and thus define infinitely many various ``Wilson loops''. It is an open question
which one of them should be used in the gauge fields\,--\,strings correspondence.
This question is important for the full treatment of the problem. However, we
conjecture that in the WKB approximation the asymptotic behavior for all reasonable
definitions should be
\begin{equation}
\label{5}
W[C]\varpropto e^{-\sqrt{\lambda }A_{\min }[C]}\; ,
\end{equation}
 and in the same approximation
\begin{equation}
\label{6}
\hat{L}(s)W[C]\approx 0\: .
\end{equation}
 We are assuming here that we are dealing with any conformal version of gauge
theory. The coupling is strong and \( \lambda \gg 1 \). When the couplings
are running, the WKB approximation in general is not applicable.

In the second variational derivative of \( W \): 
\[
\frac{\delta ^{2}W}{\delta c_{\mu }(s)\delta c_{\mu }(s')}=\left( \lambda \frac{\delta A}{\delta c_{\mu }(s)}\frac{\delta A}{\delta c_{\mu }(s')}-\sqrt{\lambda }\frac{\delta ^{2}A}{\delta c_{\mu }(s)\delta c_{\mu }(s')}\right) W\; ,\]
 the first term in brackets has no singularity for \( s\to s' \) and does not
contribute to \( \hat{L}(s) \).\footnote{%
This fact was considered in \cite{6} as a check that the loop equation is satisfied.
Notice however that in this order an arbitrary functional \( A \) will pass
this check. 
} Thus, the check that string theory in AdS space satisfies the loop equations
of motion of a gauge theory reduces to the problem of calculating
\[
\hat{L}(s)A_{\min }[c(s)]=?\]
 Solving this problem is the main objective of the present work.

\section{String Lagrangians}

Let us begin with setting up the general framework (see also \cite{1}). The
minimal area can be described with the use of the Dirichlet functional
\begin{equation}
\label{5*}
S=\frac{1}{2}\int _{\mathcal{D}}d^{2}\xi \, G_{MN}\bigl (x(\xi )\bigr )\partial _{a}x^{M}(\xi )\partial _{a}x^{N}(\xi )=\frac{1}{2}\int _{\mathcal{D}}\frac{d^{2}\xi }{y^{2}}\left[ \left( \partial _{a}\bx \right) ^{2}+\left( \partial _{a}y\right) ^{2}\right] \: ,
\end{equation}
 where \( x^{M}=(y,\bx )=(y,x^{m}) \) are the coordinates in the \( (D+1) \)-dimensional
AdS space with the metric \( G_{MN}=y^{-2}\delta _{MN} \) (\( M=0,\ldots ,D \),
\( m=1,\ldots ,D \)). This functional must be minimized with the boundary conditions
\begin{equation}
\label{6*}
\bx |_{\partial \mathcal{D}}=\bc \bigl (\alpha (s)\bigr ),\qquad y|_{\partial \mathcal{D}}=0\: ,
\end{equation}
 where the reparametrization \( \alpha (s) \) must be chosen so that the classical
action
\begin{equation}
\label{6**}
F\left[ \bc \bigl (\alpha (s)\bigr )\right] =\min _{\{\bx ,y\}}S[\bx (\xi ),y(\xi )]
\end{equation}
 is stationary. That gives the minimal area
\begin{equation}
\label{7}
A[\bc (s)]=\min _{\{\alpha (s)\}}F\left[ \bc \bigl (\alpha (s)\bigr )\right] \: .
\end{equation}
 Minimization (\ref{7}) is equivalent to the Virasoro constraints imposed on
the classical solution
\begin{equation}
\label{8}
\left\{ \begin{array}{l}
\dis T_{\perp \perp }=\frac{1}{y^{2}}\left[ \left( \partial _{\tau }\bx \right) ^{2}+\left( \partial _{\tau }y\right) ^{2}-\left( \partial _{\sigma }\bx \right) ^{2}-\left( \partial _{\sigma }y\right) ^{2}\right] =0\: ,\smallskip \\
\dis T_{\perp \Vert }=\frac{1}{y^{2}}\left[ \partial _{\tau }\bx \partial _{\sigma }\bx +\partial _{\tau }y\partial _{\sigma }y\right] =0\: ,
\end{array}\right. 
\end{equation}
 where \( \tau =\xi _{0} \), \( \sigma =\xi _{1} \). Our strategy will be
to calculate the second variation of \( F[\bc (s)] \) and to derive its short-distance
expansion as \( s\to s' \). After that we will translate the result to \( A[\bc (s)] \). 

The equations of motion for (\ref{5*}) have the form
\begin{equation}
\label{9}
\left\{ \begin{array}{l}
\dis \partial _{a}\left( \frac{1}{y^{2}}\partial _{a}\bx \right) =0\; ,\\
\dis \partial ^{2}y=\frac{1}{y}\left[ (\partial _{a}y)^{2}-\left( \partial _{a}\bx \right) ^{2}\right] \; .
\end{array}\right. 
\end{equation}
 The boundary conditions are
\[
\left\{ \begin{array}{l}
\bx (0,\sigma )=\bc (\sigma )\; ,\\
y(0,\sigma )=0\; .
\end{array}\right. \]
 It is straightforward to check that the small \( \tau  \) expansion of the
solution has the form (see \cite{1})
\begin{equation}
\label{10}
\left\{ \begin{array}{l}
\dis \bx (\tau ,\sigma )=\bc (\sigma )+\frac{1}{2}\bff (\sigma )\tau ^{2}+\frac{1}{3}\bg (\sigma )\tau ^{3}+\ldots \\
\dis y(\tau ,\sigma )=a(\sigma )\tau +\frac{1}{3}b(\sigma )\tau ^{3}+\ldots 
\end{array}\right. 
\end{equation}
 Direct substitution of (\ref{10}) into (\ref{9}) gives the relations
\begin{equation}
\label{11}
\left\{ \begin{array}{l}
\dis a^{2}(\sigma )=\left( \bc '(\sigma )\right) ^{2}\; ,\\
\dis \bff =\left( \bc '\right) ^{2}\frac{d}{d\sigma }\left( \frac{\bc '}{\left( \bc '\right) ^{2}}\right) \; ,
\end{array}\right. 
\end{equation}
 while the functions \( b \) and \( \bg  \) remain arbitrary. They must be
determined from the global considerations. After this is done, all higher terms
involved in the expansion (\ref{10}) are uniquely determined. For example the
term \( y\sim (1/4)e(\sigma )\tau ^{4} \) will be given by
\begin{equation}
\label{12}
e(\sigma )=-\frac{2}{a}(\bff \bg )-\frac{2}{3a}(\bc '\bg ')\: ,
\end{equation}
 etc.

The stress tensors (\ref{8}) at the boundary are also easily determined. Substituting
(\ref{10}) into (\ref{8}) we obtain
\begin{equation}
\label{13}
\left\{ \begin{array}{l}
\dis T_{\perp \perp }|_{\tau \to 0}=\frac{1}{a^{2}}\left[ \bff ^{2}-(\bc '\bff ')-(a')^{2}+2ab\right] \: ,\smallskip \\
\dis T_{\perp \Vert }|_{\tau \to 0}=\frac{1}{a^{2}}(\bc '\bg )\: .
\end{array}\right. 
\end{equation}
 Notice that the terms \( \sim 1/\tau ^{2} \) in \( T \) cancel automatically
due to (\ref{11}). This is because \( T \) is an analytic function which cannot
have such a singularity.

One more relation which we need is the expression for the variational derivative
\( \delta F/\delta \bc (s) \). From (\ref{5*}) by the standard integration
by parts we obtain
\begin{eqnarray}
\delta F & = & -\int \frac{d\sigma }{y^{2}}\left[ \partial _{\tau }x^{m}\delta x^{m}+\partial _{\tau }y\, \delta y\right] _{\tau \to 0}\nonumber \\
 & = & -\int \frac{d\sigma }{a^{2}\tau ^{2}}\left[ (f_{m}\tau +g_{m}\tau ^{2})\delta x_{m}+(a+b\tau ^{2})\delta y\right] _{\tau \to 0}\label{14} \\
 & = & -\frac{1}{\tau }\int \left[ \frac{f_{m}}{a^{2}}\delta x_{m}+\frac{\delta a}{a}\right] -\int \frac{d\sigma }{a^{2}}g_{m}\delta x_{m}\nonumber 
\end{eqnarray}
 (we assume here that the contour remains at \( y=0 \)). The first divergent
term is zero due to (\ref{11}), which means that the divergent part of this
action is constant. We obtain
\begin{equation}
\label{15}
\frac{\delta F}{\delta \bc (\sigma )}=-\frac{\bg (\sigma )}{a^{2}}\: .
\end{equation}

\section{\label{green}Second variation of the Dirichlet functional}

In order to act with the loop operator, we must find the second variation of
\( F \). After substituting in (\ref{5*}) the expression 
\begin{equation}
\label{16}
x_{M}\Rightarrow x_{M}(\xi )+y(\xi )\psi _{M}(\xi )
\end{equation}
 and expanding to the second order, we obtain after more or less standard calculations
\begin{eqnarray}
 &  & S=S_{0}+S_{1}^{bdry}+S_{2}+S_{2}^{bdry}\textrm{ },\nonumber \\
 &  & S_{2}=\frac{1}{2}\int d^{2}\xi \left[ \left( \nabla _{\alpha }\psi _{M}\right) ^{2}+r_{MN}(\xi )\psi _{M}\psi _{N}\right] \: ,\label{17} 
\end{eqnarray}
 where 
\begin{eqnarray}
\nabla _{\alpha }\psi _{M} & = & \partial _{\alpha }\psi _{M}+\left( \omega _{\alpha }\right) _{MN}\psi _{N}\: ,\nonumber \\
r_{MN} & = & \frac{1}{y^{2}}\left[ \left( \partial _{\alpha }x_{K}\right) ^{2}\delta _{MN}-\partial _{\alpha }x_{M}\partial _{\alpha }x_{N}\right] \: ,\label{18} \\
\left( \omega _{\alpha }\right) _{MN} & = & \frac{1}{y}\left[ \partial _{\alpha }x_{N}\delta _{M0}-\partial _{\alpha }x_{M}\delta _{N0}\right] \: .\nonumber 
\end{eqnarray}
 The terms \( S_{1,2}^{bdry} \) containing the first and second order boundary
contributions are also easily calculated, but we do not need them below. 

The second variation of the action \( S \) with respect to the \( \bc (\sigma ) \)
will be obtained if we find the classical solution of the linearized problem
with the action \( S_{2} \) 
\begin{equation}
\label{19}
-\nabla ^{2}\psi _{M}+r_{MN}\psi _{N}=0\: 
\end{equation}
 with the boundary condition
\begin{equation}
\label{20}
\left\{ \begin{array}{l}
\dis \psi _{m}|_{\tau \to 0}\to \frac{\delta c_{m}(\sigma )}{y}=\frac{\delta c_{m}(\sigma )}{a(\sigma )\tau }\: ,\\
\psi _{0}|_{\tau \to 0}=O(1)\: .
\end{array}\right. 
\end{equation}
 When this is done, we will expand \( \psi _{m}(\tau ,\sigma ) \) up to the
term \( \propto \tau ^{2} \), which will give us the variation of the \( \bg  \)-factor
and, according to (\ref{15}), the second variation of \( F \). 

Another, equivalent strategy (which requires knowing \( S_{2}^{bdry} \)) would
be to substitute this solution in (\ref{17}) and find the kernel of the resulting
quadratic functional. The standard integration by parts gives in this case
\begin{eqnarray}
\delta ^{(2)}S & = & \lim _{\tau \to 0}\left[ -\frac{1}{2}\int \psi _{M}(\tau ,\sigma )\nabla _{\tau }\psi _{M}(\tau ,\sigma )\, d\sigma +S_{2}^{bdry}\right] \nonumber \\
 & = & \frac{1}{2}\int \varkappa _{mn}(\sigma ,\sigma ')\delta c_{m}(\sigma )\delta c_{n}(\sigma ')\, d\sigma \, d\sigma '.\label{21} 
\end{eqnarray}
 Here the kernel \( \varkappa (\sigma ,\sigma ') \) can be expressed through
the Green function of the equation (\ref{19}).

This Green function is not known for a general classical solution \( \bx =\bx (\xi ) \),
\( y=y(\xi ) \). Fortunately, all we need for our task is its short distance
expansion as \( \sigma \to \sigma ' \). And this is relatively easy to obtain.
Let us begin with the leading singularity. Substitution of (\ref{10}) into
(\ref{18}) gives in the highest order
\begin{eqnarray}
 &  & \left( \omega _{1}\right) _{0m}\approx \frac{c_{m}'}{a\tau }\: ,\nonumber \\
 &  & r_{mn}\approx \frac{1}{\tau ^{2}}\left[ 2\delta _{mn}-\frac{c_{m}'c_{n}'}{a^{2}}\right] ,\qquad r_{00}\approx \frac{1}{\tau ^{2}}\: .\label{22} 
\end{eqnarray}
 Substituting this into (\ref{17}), we obtain in this approximation 
\begin{equation}
\label{23}
S_{2}=\frac{1}{2}\int d^{2}\xi \left[ \left( \partial _{\tau }\psi _{M}\right) ^{2}+\left( \partial _{\sigma }\psi _{M}\right) ^{2}+\frac{2}{\tau ^{2}}(\psi _{M})^{2}+\frac{c_{m}'}{\tau a}\left( \psi _{m}\partial _{\sigma }\psi _{0}-\psi _{0}\partial _{\sigma }\psi _{m}\right) \right] \: .
\end{equation}
 We kept the terms \( \sim 1/\tau ^{2} \) in the Lagrangian. As we will see
in a moment, this is sufficient in the leading order. In this order, we can
also neglect the \( \sigma  \)-dependence of \( c_{m}' \). Then the action
splits into two parts
\begin{eqnarray}
 &  & S_{2}=S^{\perp }+S^{\Vert }\: ,\nonumber \\
 &  & S^{\perp }=\frac{1}{2}\int d^{2}\xi \left[ \left( \partial _{\alpha }\psi _{i}\right) ^{2}+\frac{2}{\tau ^{2}}\left( \psi _{i}\right) ^{2}\right] \: ,\label{24} \\
 &  & S^{\Vert }=\frac{1}{2}\int d^{2}\xi \left[ \left( \partial _{\alpha }\psi _{0}\right) ^{2}+\left( \partial _{\alpha }\psi _{1}\right) ^{2}+\frac{2}{\tau ^{2}}\left( \psi _{0}^{2}+\psi _{1}^{2}\right) +\frac{1}{\tau }\left( \psi _{1}\partial _{\sigma }\psi _{0}-\psi _{0}\partial _{\sigma }\psi _{1}\right) \right] \: ,\nonumber 
\end{eqnarray}
 where \( i=2,\ldots ,D \), and we chose the contour to run in the \( x_{1} \)
direction. 

Let us now solve the Dirichlet problem for the action (\ref{24}). In the case
of \( S^{\perp } \) we get
\begin{eqnarray}
 &  & \left( \partial _{\tau }^{2}-p^{2}\right) \psi _{i}-\frac{2}{\tau ^{2}}\psi _{i}=0\: ,\nonumber \\
 &  & \psi _{i}(p,\tau )=\tau ^{-1}(1+|p|\tau )e^{-|p|\tau }\delta c_{i}(p)\: ,\label{25} 
\end{eqnarray}
 where we introduced the Fourier transformed quantities \( \psi _{i}(p,\tau ) \)
instead of \( \psi _{i}(\sigma ,\tau ) \). 

The longitudinal case is slightly more complex, but complex variables solve
it. If we introduce \( \Psi =\psi _{1}+i\psi _{0} \), we get
\begin{equation}
\label{26}
\left( \partial _{\tau }^{2}-p^{2}\right) \Psi -\frac{2}{\tau ^{2}}\Psi =\frac{2}{\tau }p\Psi \: .
\end{equation}
 The solution of this equation takes the form
\begin{eqnarray}
 &  & \Psi (p,\tau )=\tau ^{-1}e^{-|p|\tau }\left[ 1+\theta (-p)\left( 2|p|\tau +2p^{2}\tau ^{2}\right) \right] \delta c_{1}(p)\: ,\nonumber \\
 &  & \left\{ \begin{array}{l}
\dis \psi _{1}(p,\tau )=\frac{1}{2}\left[ \Psi (p,\tau )+\Psi ^{*}(-p,\tau )\right] \: ,\smallskip \\
\dis \psi _{0}(p,\tau )=\frac{1}{2i}\left[ \Psi (p,\tau )-\Psi ^{*}(-p,\tau )\right] \: .
\end{array}\right. \label{27} 
\end{eqnarray}
 As we discussed above, all we need from these solutions are the \( \bg  \)-factors,
that is the coefficients in front of \( \tau ^{2} \) as \( \tau \to 0 \).

Expanding expressions (\ref{25}) and (\ref{27}), we get
\begin{equation}
\label{28}
\left\{ \begin{array}{l}
g_{i}(p)=|p|^{3}\delta c_{i}(p)\: ,\\
g_{1}(p)=-2|p|^{3}\delta c_{1}(p)\: .
\end{array}\right. 
\end{equation}
 The functional (\ref{21}) can be written in a convenient form if we introduce
the Fourier transform
\begin{equation}
\label{29}
\varkappa (\sigma ,\sigma ')=\int \tilde{\varkappa }\left( \frac{\sigma +\sigma '}{2},p\right) e^{ip(\sigma -\sigma ')}dp\: .
\end{equation}
 From (\ref{28}) and (\ref{15}) it follows that
\begin{equation}
\label{30}
\tilde{\varkappa }_{mn}(p,\sigma )\mathop {=}_{p\to \infty }\frac{|p|^{3}}{\bc '(\sigma )^{2}}\left[ 2\frac{c_{m}'c_{n}'}{(\bc ')^{2}}-\left( \delta _{mn}-\frac{c_{m}'c_{n}'}{(\bc ')^{2}}\right) \right] \: ,
\end{equation}
 where we restored the general form of \( c_{m} \).

Let us now discuss the range of validity of this formula. In its derivation
we neglected many terms in (\ref{19}). There are terms containing higher powers
of \( \tau  \) coming from higher order terms in (\ref{10}) (we kept only
the first ones). Also, we neglected the \( \sigma  \)-dependence of \( c_{m}'(\sigma ) \)
in (\ref{23}). 

This is legitimate, since the values of \( \tau  \) involved in our calculations
are \( \tau \lesssim 1/|p| \). Hence, if we treat the \( \tau  \)-correction
as a perturbation in (\ref{23}), we will obtain a contribution to \( \varkappa (p,\sigma ) \)
suppressed by \( 1/|p| \). The same is true for the \( \sigma  \)-correction
of \( c_{m}'(\sigma ) \)---if we expand \( c_{m}'(\sigma )=c_{m}'(\sigma _{0})+c_{m}''(\sigma )(\sigma -\sigma _{0}) \),
then in perturbation theory \( \sigma -\sigma _{0}\Rightarrow i\partial /\partial p \)
in the \( (p,\sigma ) \)-representation. So, our conclusion is that (\ref{30})
is indeed the leading singularity, and we need \( |p|\gg |\bc ''|/|\bc '| \)
for it to be valid.

\section{\label{diman}Dimensional analysis}

In the previous section, we found the leading singularity \( \sim |p|^{3} \)
in \( \tilde{\varkappa }_{mn}(p,\sigma ) \). However, this is just the beginning
of the story, since to calculate the action of the loop operator we need to
pick up terms \( \sim p^{0} \) in \( \tilde{\varkappa }_{mm}(p,\sigma ) \).
The general structure has the form
\begin{equation}
\label{31}
\tilde{\varkappa }_{mm}(p,\sigma )\mathop {=}_{p\to \infty }(3-D)\frac{|p|^{3}}{\bigl (\bc '(\sigma )\bigl )^{2}}+A_{1}(\sigma )p^{2}+A_{2}(\sigma )|p|+A_{3}(\sigma )+\ldots 
\end{equation}
 According to our analysis, in order to calculate \( A_{1} \) we have to expand
\( r_{MN} \) up to \( 1/\tau  \) and \( \omega _{\alpha } \) up to \( \tau ^{0} \).
\( A_{2} \) and \( A_{3} \) will require further expansion. We will also have
to expand \( \bc '(\sigma ) \) to the needed order.

Direct use of the perturbation theory is straightforward because we know explicit
Green functions for the unperturbed equations (\ref{25}) and (\ref{26}), but
cumbersome (see Appendix).

We can greatly simplify our task by noticing that the functions \( A_{1,2,3}(\sigma ) \)
\emph{depend locally on the properties of the coefficients \( r_{MN} \) and
\( \omega _{\alpha } \) in} (\ref{19}) \emph{in the limit \( \tau \to 0 \)}.
That means that they locally depend on the quantities \( \bc ,\bff ,\bg ,a,b,\ldots  \)
appearing in the expansion (\ref{10}). Moreover, we can uncover this dependence
by the \emph{simple dimensional analysis.}

The rules are as follows. We already know that
\begin{equation}
\label{31*}
\tau \sim \sigma \sim 1/p\: ,
\end{equation}
 where by \( \sigma  \) we mean deviation from the middle point. Then
\begin{eqnarray}
 &  & \bc '(\sigma )\sim \tau ^{0}\: ,\quad \bc ''(\sigma )\sim 1/\tau \: ,\quad \bc '''(\sigma )\sim 1/\tau ^{2}\: ,\nonumber \\
 &  & \bff \sim 1/\tau \: ,\quad \bg \sim 1/\tau ^{2}\: ,\quad a\sim \tau ^{0}\: ,\quad b\sim 1/\tau ^{2}\: .\label{32} 
\end{eqnarray}
 Notice that with these assignments it follows from (\ref{13}) that \( T_{\perp \perp }\sim T_{\perp \Vert }\sim 1/\tau ^{2} \),
as it should be.

Now, from (\ref{31}) the scaling of \( A_{1,2,3} \) must be the following:
\begin{equation}
\label{33}
A_{1}\sim p\sim 1/\tau \, ,\quad A_{2}\sim p^{2}\sim 1/\tau ^{2}\, ,\quad A_{3}\sim 1/\tau ^{3}\: .
\end{equation}
 Let us start now constructing these quantities. For \( A_{1} \) the only thinkable
combination is
\begin{equation}
\label{34}
A_{1}=f(\bcp ^{2})(\bc '\bc '').
\end{equation}
 It follows that we must have \( A_{1}=0 \), because the expression (\ref{34})
is odd under the change \( \sigma \to -\sigma  \), while the equations (\ref{19})
and (\ref{21}) preserve this parity. Hence we confirm the result of \cite{1}
that there is no term \( \propto \delta ''(\sigma _{1}-\sigma _{2}) \) in \( \delta ^{2}A/\delta \bc (\sigma _{1})\delta \bc (\sigma _{2}) \)
(this term would break the zigzag symmetry).

For \( A_{2} \) we have
\begin{equation}
\label{35}
A_{2}=\alpha _{1}(\bc '')^{2}+\alpha _{2}(\bc '\bc '')^{2}+\alpha _{3}(\bc '\bc ''')+\alpha _{4}b+\alpha _{5}(\bc '\bg )\: .
\end{equation}
 Notice that \( \alpha _{5}=0 \) because \( \bg  \) is even under \( \sigma \to -\sigma  \)
while \( \bc ' \) is odd. If the Virasoro conditions are satisfied, we can
express \( b \) in terms of the contour from (\ref{13}). The quantities \( \alpha _{k} \)
are functions of \( (\bc ')^{2} \). It is easy to find them using the symmetry
\( \bc \to \lambda \bc  \). Being a second derivative, \( \varkappa  \) must
scale as \( c^{-2} \). Hence we conclude that 
\begin{equation}
\label{35*}
\alpha _{1}\propto (\bc ')^{-4}\: ,\quad \alpha _{2}\propto (\bc ')^{-6}\: ,\quad \alpha _{3}\propto (\bc ')^{-4}\: ,\quad \alpha _{4}\propto |\bc '|^{-3}.
\end{equation}
 To calculate the numerical coefficients, one may use the wavy line limit of
\cite{1}. We will not attempt it here. 

Now we come to the most interesting term \( A_{3} \), representing the action
of the loop operator. We have the following possible structures:
\begin{equation}
\label{36}
A_{3}=\beta _{1}\frac{(\bc ''\bg )}{(\bc ')^{4}}+\beta _{2}\frac{(\bc '\bc '')(\bc '\bg )}{(\bc ')^{6}}+\beta _{3}\frac{(\bc '\bg )'}{(\bc ')^{4}}\: .
\end{equation}
 Once again dimensionally possible terms \( (\bc '\bc ^{(4)}) \), \( (\bc ''\bc ''') \),
\( (\bc '')^{2}(\bc '\bc '') \), \( (\bc '\bc ''')(\bc '\bc '') \), \( b' \)
and \( b(\bc '\bc '') \) are forbidden by parity. The main novelty here is
the appearance of \( \bg  \). This coefficient depends on \( \bc (\sigma ) \)
non-locally. However, we can express it from (\ref{15}) as 
\begin{equation}
\label{37}
\bg =-(\bc ')^{2}\left( \frac{\delta F}{\delta \bc (\sigma )}\right) \: .
\end{equation}
 Thus (\ref{36}) gives a remarkable relation between the loop operator and
the first derivative of \( F \):
\begin{equation}
\label{38}
\hat{L}(\sigma )F=-\frac{\beta _{1}}{(\bc ')^{2}}\left( \bc ''\frac{\delta F}{\delta \bc }\right) +\ldots 
\end{equation}
 Since the remaining terms contain the combination 
\begin{equation}
\label{39}
(\bc '\bg )=(\bc ')^{2}T_{\perp \Vert }\: ,
\end{equation}
 they drop out if the Virasoro conditions are satisfied.

\section{Second variation of the minimal area}

In order to calculate the minimal area, we have to use the relation (\ref{7}).
By taking \( \alpha (s)\Rightarrow s+\alpha (s) \) and \( c_{m}(s)\Rightarrow c_{m}(s)+\delta c_{m}(s) \),
and by expanding (\ref{7}) to the second order, we obtain
\begin{eqnarray}
\delta ^{(2)}A & = & \min _{\left\{ \alpha (s)\right\} }\Biggl [\int \frac{\delta F}{\delta c_{m}(s)}\left( \delta c_{m}'(s)\alpha (s)+\frac{1}{2}c_{m}''(s)\alpha ^{2}(s)\right) \, ds\nonumber \\
 &  & +\frac{1}{2}\int \frac{\delta ^{2}F}{\delta c_{m}(s)\delta c_{n}(s')}\Bigl (\delta c_{m}(s)+\alpha (s)c_{m}'(s)\Bigr )\Bigl (\delta c_{n}(s')+\alpha (s')c_{n}'(s')\Bigr )\, ds\, ds'\Biggr ]\: .\label{40} 
\end{eqnarray}
 In principle this formula solves the problem of expressing the second derivative
of the minimal area \( A \) through the Dirichlet functional \( F \). For
that we have to exclude \( \alpha (s) \) from it by the use of extremality
condition. Again this can be done by the use of the short distance expansions
of the kernels involved. However, for our limited purposes in this paper we
do not have to do it, since the dimensional analysis for \( \hat{L}A \) is
the same as for \( \hat{L}F \). 

We will discuss in Appendix direct calculation of \( \hat{L}A \), while here
we can just conclude from (\ref{36}) that
\begin{equation}
\label{41}
\hat{L}A[\bc (s)]=\gamma \frac{(\bc ''\bg )}{(\bc ')^{4}}=-\frac{\gamma }{(\bc ')^{2}}\left( \bc ''(s)\frac{\delta A[\bc (s)]}{\delta \bc (s)}\right) \: .
\end{equation}
 To calculate the constant \( \gamma  \), it is sufficient to consider the
wavy line limit of \cite{1}. In this case we already obtained
\begin{equation}
\label{42}
\gamma =D-4\: .
\end{equation}
 We can also understand why \( \gamma \propto D-4 \) without any calculations.
Namely, at \( D=4 \) the \( \hat{L} \) operator is conformally invariant,
which means that not only \( A[\bc (s)] \) is invariant under conformal transformations
but also \( \hat{L}A[\bc (s)] \).\footnote{%
This property of \( \hat{L} \) is to be expected, since the Yang-Mills equations
are conformally invariant exactly at \( D=4 \). For the direct (and not completely
trivial) proof see \cite{1}.
} However these transformations can be used to set \( \bc ''(\sigma )=0 \) (a
familiar example---we can make straight lines out of circles). Thus the only
way in which (\ref{41}) may be consistent with conformal symmetry is \( \gamma =0 \)
at \( D=4 \). As it was shown in \cite{1}, if \( D\ne 4 \) there is an inhomogeneous
term in the conformal transformation of \( \hat{L}A \) proportional to \( (D-4) \).
That explains the origin of (\ref{42}). 

So, the formulae (\ref{41}) and (\ref{42}) solve the main problem addressed
in this paper---verification of the loop equation in the WKB limit. This is
a first step in deriving the loop equation for the full string theory functional
integral. This problem requires much more concrete expressions both for the
Wilson loop and for the string Lagrangians. We hope to address it in the future.
Here we will just point out a general mechanism by which the non-linear terms
in (\ref{2}) can be generated. The equation (\ref{38}) tells us that \( \hat{L}(\sigma )F\propto T_{\perp \Vert }(\sigma ) \).
In the classical theory \( T_{\perp \Vert }=0. \) However in quantum theory
expectation values \( \langle T_{\perp \Vert }\rangle  \) are dominated by
the pinched disc \cite{4}. This must be the source of the non-linearity in
the loop equations. However there are still some technical difficulties in implementing
this idea.

\section*{Acknowledgements}

The work of A.P. was partially supported by NSF grant PHY-98-02484.

\appendix\setcounter{section}{1}\setcounter{equation}{0}

\section*{Appendix: Calculation of \protect\( \hat{L}A\protect \)}

In this appendix we give a direct calculation of \( \hat{L}A \) along the lines
of Section \ref{green}. This will provide an independent check of (\ref{42}),
and also to some extent justify and exemplify the dimensional analysis method
of Section \ref{diman}. 

It will be convenient to switch from the conformal coordinates \( (\sigma ,\tau ) \)
to the static gauge
\begin{equation}
\label{c1}
x_{0}(t,s)=t\: .
\end{equation}
 The minimal surface is then given by
\begin{equation}
\label{c2}
\bx (t,s)=\bc (s)+\frac{1}{2}\bff (s)t^{2}+\frac{1}{3}\bg (s)t^{3}+\frac{1}{4}\bh (s)t^{4}+\ldots 
\end{equation}
 A derivation similar to (\ref{14}) shows that in this gauge
\begin{equation}
\label{c3}
\frac{\delta A}{\delta \bg (s)}=\sqrt{\bcp ^{2}}\bg (s)\: .
\end{equation}
 To simplify further calculations, we choose the coordinate \( s \) on the
\emph{unperturbed} world sheet so that
\begin{equation}
\label{c4}
\left\{ \begin{array}{ll}
\bcp ^{2}(s)\equiv 1 & (\Rightarrow \: \bff =\bc '')\: ,\\
\bigl (\dot{\bx }(t,s)\bc '(s)\bigl )\equiv 0 & (\Rightarrow \: (\bff \bc ')=(\bg \bc ')=(\bh \bc ')=\cdots =0)\: .
\end{array}\right. 
\end{equation}
 Let us choose a tangent space basis \( \{\bt ,\bn ^{a}\} \), \( a=1,\ldots ,D \),
at every point of the curve \( \bc (s) \) so that
\begin{equation}
\label{c5}
\bt =\frac{\bc '}{\sqrt{\bcp ^{2}}}\: ,\quad (\bn ^{a}\bt )=0\: ,\quad (\bn ^{a}\bn ^{b})=\delta ^{ab}\: .
\end{equation}
 Then every variation \( \delta \bc  \) can be decomposed into the normal and
longitudinal part:
\begin{equation}
\label{c6}
\delta \bc =\bn ^{a}(\bn ^{a}\delta \bc )+\bt (\bt \, \delta \bc )=\delta \bc ^{\perp }+\delta \bc ^{\Vert }\: .
\end{equation}
 The second variation of \( A \) takes the form
\begin{eqnarray}
\delta ^{(2)}A & = & \frac{1}{2}\int \! \! \! \int \frac{\delta ^{2}A}{\delta c_{i}(s)\delta c_{k}(s')}\delta c_{i}(s)\delta c_{k}(s')\nonumber \\
 & = & \frac{1}{2}\int \! \! \! \int \frac{\delta ^{2}A}{\delta \bn ^{a}(s)\delta \bn ^{b}(s')}\bigl (\bn ^{a}\delta \bc (s)\bigl )\bigl (\bn ^{b}\delta \bc (s')\bigl )\label{c7} \\
 &  & \quad +\, 2\frac{\delta ^{2}A}{\delta \bt (s)\delta \bn ^{a}(s')}\bigl (\bt \, \delta \bc (s)\bigl )\bigl (\bn ^{a}\delta \bc (s')\bigl )\, +\, \frac{\delta ^{2}A}{\delta \bt (s)\delta \bt (s')}\bigl (\bt \, \delta \bc (s)\bigl )\bigl (\bt \, \delta \bc (s')\bigl )\: ,\nonumber 
\end{eqnarray}
 where 
\begin{eqnarray}
\frac{\delta ^{2}A}{\delta \bn ^{a}(s)\delta \bn ^{b}(s')} & = & \frac{\delta ^{2}A}{\delta c_{i}(s)\delta c_{k}(s')}\: n^{a}_{i}(s)n^{b}_{k}(s')\: ,\nonumber \\
\frac{\delta ^{2}A}{\delta \bt (s)\delta \bn ^{a}(s')} & = & \frac{\delta ^{2}A}{\delta c_{i}(s)\delta c_{k}(s')}\: t_{i}(s)n^{a}_{k}(s')\: ,\label{c8} \\
\frac{\delta ^{2}A}{\delta \bt (s)\delta \bt (s')} & = & \frac{\delta ^{2}A}{\delta c_{i}(s)\delta c_{k}(s')}\: t_{i}(s)t_{k}(s')\: .\nonumber 
\end{eqnarray}
 Differentiating the reparametrization invariance condition \( t_{i}(s)\cdot \delta A/\delta c_{i}(s)=0 \)
gives
\begin{equation}
\label{c9}
\frac{\delta ^{2}A}{\delta c_{i}(s)\delta c_{k}(s')}\: t_{i}(s)=-\frac{\delta A}{\delta c_{i}(s)}\cdot \frac{\delta t_{i}(s)}{\delta c_{k}(s')}\: .
\end{equation}
 Now it follows from (\ref{c7})-(\ref{c9}) that 
\begin{eqnarray}
\frac{\delta ^{2}A}{\delta c_{i}(s)\delta c_{i}(s')} & = & \frac{\delta ^{2}A}{\delta \bn ^{a}(s)\delta \bn ^{b}(s')}\bigl (\bn ^{a}(s)\bn ^{b}(s')\bigl )\nonumber \\
 &  & \quad -\frac{\delta A}{\delta c_{i}(s)}\cdot \frac{\delta t_{i}(s)}{\delta c_{k}(s')}\left[ 2n_{k}^{a}(s')\bigl (\bt (s)\bn ^{a}(s')\bigl )+t_{k}(s')\bigl (\bt (s)\bt (s')\bigl )\right] \nonumber \\
 & = & \frac{\delta ^{2}A}{\delta \bn ^{a}(s)\delta \bn ^{b}(s')}\bigl (\bn ^{a}(s)\bn ^{b}(s')\bigl )+\left( \frac{\delta A}{\delta \bc (s)}\bc ''(s)\right) \delta (s-s')\: ,\label{c10} 
\end{eqnarray}
 where in the last line we used
\begin{equation}
\label{c11}
\frac{\delta t_{i}(s)}{\delta c_{k}(s')}=\left[ \delta _{ik}-c_{i}'(s)c_{j}'(s)\right] \delta '(s-s')\: .
\end{equation}
 To find \( \delta ^{2}A/\delta \bn ^{a}(s)\delta \bn ^{b}(s') \), we vary
(\ref{c3}) for the second time with a normal variation \( \delta \bc =\delta \bc ^{\perp } \).
The result is
\begin{equation}
\label{c13}
\delta ^{(2)}_{\perp }A=\frac{1}{2}\int \left[ (\bg \delta \bc ^{\perp })(\bc ''\delta \bc ^{\perp })-(\delta \bg \delta \bc ^{\perp })\right] \, ds\: .
\end{equation}
 From this we see that
\begin{equation}
\label{c14}
\frac{\delta ^{2}A}{\delta \bn ^{a}(s)\delta \bn ^{a}(s')}=-\frac{\delta g^{a}(s)}{\delta \bn ^{a}(s')}+(\bg \bc '')\delta (s-s')\: ,\qquad \delta g^{a}:=(\bn ^{a}\delta \bg )\: .
\end{equation}
 It follows from this and (\ref{c10}) that
\begin{equation}
\label{c15}
\hat{L}(s)A=\textrm{coefficient before }\delta (s-s')\textrm{ in }\left( -\frac{\delta g^{a}(s)}{\delta \bn ^{a}(s')}\right) \: .
\end{equation}
 Thus we reduced the problem to studying how \( \bg (s) \) changes under normal
variations.

It is well known (a clear-cut derivation can be found in \cite{5}) that under
a normal variation \( x_{M}\Rightarrow x_{M}(t,s)+\psi _{M}(t,s) \) the area
of a minimal surface in curved space changes to the second order in \( \psi  \)
by 
\begin{equation}
\label{4.1}
S_{2}=\int |\nabla ^{\perp }\psi |^{2}-\sum _{i,j}|\langle \psi ,B(E_{i},E_{j})\rangle |^{2}-\langle R(E_{i},\psi )\psi ,E_{i}\rangle 
\end{equation}
 (modulo boundary terms). In this coordinate-free notation \( \nabla ^{\perp } \)
is the covariant derivative in the normal bundle; \( \left\{ E_{i}\right\}  \)
is an orthonormal basis of the surface tangent space; \( \langle \cdot ,\cdot \rangle  \)
is the metric of the ambient space; \( B(E_{i},E_{j}) \) is the second fundamental
form of the surface; \( R \) is the Riemann curvature tensor; integration is
taken with respect to the induced metric volume form.

Let us specify (\ref{4.1}) to our case of the surface (\ref{c1}), (\ref{c2})
in the AdS space. We assume that at every point \( (t,s) \) of the surface
we have a basis \( \left\{ n^{a}(t,s)\right\}  \) of \( (D-1) \) normal vectors
such that
\begin{equation}
\label{c16}
n^{a}(0,s)=\left( 0\atop \bn ^{a}(s)\right) ,\quad \quad n^{a}_{M}n_{M}^{b}=\delta ^{ab}\: ,
\end{equation}
 with \( \bn ^{a} \) from above. We also put \( \nu _{M}^{a}(t,s)=t\, n_{M}^{a}(t,s) \),
so that \( \nu ^{a} \) are AdS-normalized. Now \( \psi  \) can be written
as 
\begin{equation}
\label{c17}
\psi _{M}=\psi ^{a}(t,s)\nu _{M}^{a}(t,s)\: .
\end{equation}

The curvature term in (\ref{4.1}) is simple:
\begin{eqnarray}
 &  & R_{[MK][NL]}=-(G_{MN}G_{KL}-G_{ML}G_{KN})\: ,\nonumber \\
 &  & \langle R(E_{i},\psi )\psi ,E_{i}\rangle =-\langle \psi ,\psi \rangle \langle E_{i},E_{i}\rangle +\langle \psi ,E_{i}\rangle \langle \psi ,E_{i}\rangle =-2\psi ^{a}\psi ^{a}\: .\label{c18} 
\end{eqnarray}

For the kinetic term we have
\begin{eqnarray}
 &  & \int |\nabla ^{\perp }\psi |^{2}=\int d^{2}\xi \sqrt{g}g^{\alpha \beta }\nabla _{\alpha }\psi ^{a}\nabla _{\beta }\psi ^{a}\: ,\nonumber \\
 &  & \nabla _{\alpha }\psi ^{a}=\partial _{\alpha }\psi ^{a}+(w_{\alpha })^{[ab]}\psi ^{b}\: .\label{c20} 
\end{eqnarray}
 Here \( g_{\alpha \beta }=\lr {\partial _{\alpha }x,\partial _{\beta }x} \)
is the induced metric, which is found to be 
\begin{eqnarray}
g_{\alpha \beta } & = & \frac{1}{t^{2}}\left( \begin{array}{cc}
1+\bff ^{2}t^{2}+2(\bff \bg )t^{3} & *\\
\frac{1}{2}(\bff \bff ')t^{3} & 1-\bff ^{2}t^{2}-\frac{2}{3}(\bff \bg )t^{3}
\end{array}\right) +O(t^{2})\: ,\nonumber \\
\sqrt{g} & = & \frac{1}{t^{2}}\bigl (1+\frac{2}{3}(\bff \bg )t^{3}\bigl )+O(t^{2})\: .\label{c19} 
\end{eqnarray}
 The spin connection coefficients \( w_{\alpha } \) are antisymmetric and given
by
\begin{equation}
\label{c21}
(w_{\alpha })^{[ab]}=\lr {\nabla _{\alpha }\nu ^{a},\nu ^{b}}=\partial _{\alpha }n_{M}^{a}\cdot n_{M}^{b}\qquad (a\ne b)\: .
\end{equation}
 To simplify them, we choose the normals so that at the boundary 
\begin{equation}
\label{2.2}
(\bn ^{a})'=-(\bn ^{a}\bc '')\bc '
\end{equation}
 (it is easy to see that this condition is compatible with (\ref{c5})). For
this choice of gauge we see from (\ref{c21}) that \( w_{\alpha }=0 \) at the
boundary, and thus for small \( t \)
\begin{equation}
\label{c22}
w_{\alpha }(t,s)=O(t)\: .
\end{equation}

Finally, the second term in (\ref{4.1}) is \( O(\psi ^{2}K^{2}) \), where
\begin{equation}
\label{c23}
K^{2}(t,s)=\lr {B(E_{i},E_{j}),B(E_{i},E_{j})}
\end{equation}
 is the extrinsic curvature of the surface squared. This quantity is connected
with the ambient space and the surface curvature tensors by the Gauss equation:
\begin{equation}
\label{c24}
K^{2}=\lr {R(E_{i},E_{j})E_{j},E_{i}}-\lr {r(E_{i},E_{j})E_{j},E_{i}}=-2-r\: ,
\end{equation}
 where \( r \) is the scalar curvature of the induced metric \( g_{\alpha \beta } \).
One can calculate from (A.20) that \( r=-2+O(t^{4}) \), and thus
\begin{equation}
\label{c25}
K^{2}=O(t^{2})\: .
\end{equation}

By using (\ref{c18}), (\ref{c20}), (\ref{c22}) and (\ref{c25}) in (\ref{4.1})
we obtain
\begin{equation}
\label{c26}
S_{2}=\int d^{2}\xi \left[ \sqrt{g}\left( g^{\alpha \beta }\partial _{\alpha }\psi ^{a}\partial _{\beta }\psi ^{a}+2g^{\alpha \beta }w^{[ab]}_{\alpha }\partial _{\beta }\psi ^{a}\psi ^{b}+2\psi ^{a}\psi ^{a}\right) +O(t^{2}\psi ^{2})\right] \: .
\end{equation}
 Varying (\ref{c26}), we obtain the equation of motion for \( \psi  \) 
\begin{equation}
\label{c27}
\partial _{\beta }\left( \sqrt{g}g^{\alpha \beta }\partial _{\alpha }\psi ^{a}\right) -2\sqrt{g}\psi ^{a}+2\sqrt{g}g^{\alpha \beta }w_{\alpha }^{[ab]}\partial _{\beta }\psi ^{b}=O(t^{2}\psi )\: .
\end{equation}

As it was explained in Section \ref{green} (this will also be evident from
the calculation below) each power of \( t \) in the equations of motion suppresses
the perturbative contribution to the Green function by a factor of \( 1/p \).
Since the highest singularity will be \( |p|^{3} \) and we are interested in
finding terms \( \sim p^{0} \), the terms \( O(t^{2}\psi ) \) in (\ref{c27})
are negligible, being 4 orders less singular than the leading term \( \sqrt{g}\psi ^{a} \). 

The third term in the LHS of (\ref{c27}) is in principle just 2 orders less
singular than the leading terms, so its contribution to \( \delta g^{a}(s)/\delta \bn ^{b}(s') \)
can contain terms \( \propto \delta '(s-s') \) and \( \delta (s-s') \). However,
this contribution will be \emph{antisymmetric} in \( a\leftrightarrow b \)
and hence is irrelevant for the subsequent substitution into (\ref{c15}). 

Thus effectively we have to care only about the first two terms in (\ref{c27}).
Taking this into account and using (\ref{c19}), we get the following equation
for \( \phi =\phi ^{a}=t\cdot \psi ^{a} \)
\begin{eqnarray}
\partial _{t}\frac{\dot{\phi }}{t^{2}}+\frac{\phi ''}{t^{2}} & = & \bigl (\bff ^{2}+\frac{4}{3}(\bff \bg )t\bigl )(\ddot{\phi }-\phi '')\nonumber \\
 &  & {}+\frac{4}{3}(\bff \bg )\dot{\phi }-\frac{3}{2}(\bff \bff ')\phi '+(\bff \bff' )t\, \dot{\phi }'+\textrm{negligible }\: .\label{c28} 
\end{eqnarray}
 The boundary conditions are 
\begin{equation}
\label{c29}
\phi (0,s)=\phi (s)\: .
\end{equation}
 The solution of the Dirichlet problem (\ref{c28}), (\ref{c29}) can be written
as 
\begin{equation}
\label{c30}
\phi (t,s)=\int K(t,s|s')\phi (s')\, ds'\: ,
\end{equation}
 where the Green function \( K(t,s|s') \) solves (\ref{c28}) with the boundary
conditions
\begin{equation}
\label{c31}
K(0,s|s')=\delta (s-s')\: .
\end{equation}

The small \( t \) expansion of \( \phi  \) can be found through \( K \) as
follows
\begin{equation}
\label{c32}
\phi (t,s)=\phi (s)+\int \left[ t\dot{K}(0,s|s')+\frac{1}{2}t^{2}\ddot{K}(0,s|s')+\ldots \right] \phi (s')\, ds'\: .
\end{equation}
 In particular, we have 
\begin{equation}
\label{c33}
\frac{\delta \! \stackrel{{\ldots }}{\phi ^{a}}\! \! (s)}{\delta \phi ^{b}(s')}=\stackrel{{\ldots }}{K}\! (0,s|s')+\textrm{antisymmetric in }a\leftrightarrow b\: .
\end{equation}

To calculate \( K \) perturbatively, we pass from the variables \( (s,s') \)
to \( (\sigma =s-s',s') \) and perform the Fourier transform in \( \sigma  \).
We have 
\begin{equation}
\label{c34}
\partial _{s}\Rightarrow \partial _{\sigma }\Rightarrow ip,\qquad s\Rightarrow \sigma +s'=s'+i\partial /\partial p.
\end{equation}
 Now equation (\ref{c28}) takes the form 
\begin{eqnarray}
\partial _{t}\frac{\dot{K}}{t^{2}}-\frac{p^{2}K}{t^{2}} & = & \bigl [\bff ^{2}+2i(\bff \bff ')\partial _{p}+\frac{4}{3}(\bff \bg )t\bigl ](\ddot{K}+p^{2}K)\nonumber \\
 &  & +\frac{4}{3}(\bff \bg )\dot{K}-\frac{3}{2}(\bff \bff ')ipK+(\bff \bff' )tip\dot{K}+\textrm{negligible },\label{4.19} 
\end{eqnarray}
 where the coefficients in the RHS are taken at \( s' \). 

The unperturbed Green function is 
\begin{equation}
\label{c36}
K_{0}=e^{-|p|t}(1+|p|t).
\end{equation}
 We are looking for \( K \) in the form of a series
\begin{equation}
\label{c37}
K(p,t|s')=K_{0}+K_{1}+K_{2}+K_{3}+\ldots \: ,
\end{equation}
where 
\begin{equation}
\label{d1}
K_{i}\sim \frac{1}{p^{i}}e^{-|p|t}\qquad (p\gg 1)\: .
\end{equation}
 The equation (\ref{4.19}) can be schematically written as 
\begin{equation}
\label{d2}
A_{0}K=A_{2}K+A_{3}K+\ldots \: ,
\end{equation}
 where we indicated the relative order of singularity of the differential operators.
It follows by simple counting that the \( K_{1} \) correction is absent from
(\ref{c37}) and that \( \tilde{K}=K_{2}+K_{3} \) is found from the equation
\begin{equation}
\label{d3}
\left\{ \begin{array}{l}
A_{0}\tilde{K}=(A_{2}+A_{3})K_{0}\: ,\\
\tilde{K}|_{t=0}=0\: .
\end{array}\right. 
\end{equation}
 Thus substituting \( K_{0} \) into the RHS of (\ref{4.19}) gives the equation
for \( \tilde{K} \) 
\begin{eqnarray}
 &  & \partial _{t}\frac{\partial _{t}\tilde{K}}{t^{2}}-\frac{p^{2}\tilde{K}}{t^{2}}=f(p,t):=2\bff ^{2}p^{2}e^{-|p|t}|p|t\nonumber \\
 &  & \qquad {}+|p|e^{-|p|t}\Bigl \{\frac{8}{3}(\bff \bg )p^{2}t^{2}-\frac{4}{3}(\bff \bg )|p|t+i\text {sign}p(\bff \bff ')\left( \frac{9}{2}|p|t-3p^{2}t^{2}-\frac{3}{2}\right) \Bigr \}\: .\label{d4} 
\end{eqnarray}
 This inhomogeneous problem has the solution 
\begin{equation}
\label{d5}
\tilde{K}(p,t)=\int G_{p}(t,\tau )f(p,\tau )\, d\tau \: 
\end{equation}
 with the Green function
\begin{equation}
\label{d6}
\left\{ \begin{array}{l}
\dis G_{p}(t_{1},t_{2})=\frac{1}{2|p|^{3}}F_{0}(|p|t_{<})F_{\infty }(|p|t_{>})\: ,\\
F_{\infty }(t)=e^{-t}(1+t)\: ,\\
F_{0}(t)=e^{t}(1-t)-F_{\infty }(t)\: .
\end{array}\right. 
\end{equation}
 Routine power expansion shows that
\begin{equation}
\label{d7}
(K_{0}+\tilde{K})(p,t)=1-\frac{1}{2}t^{2}p^{2}+\frac{1}{3}t^{3}\left( |p|^{3}-\bff ^{2}|p|-(\bff \bg )-(\bff \bff ')i\text {sign}p\right) \: .
\end{equation}
 Thus it follows from this and (\ref{c33}) that 
\begin{equation}
\label{c38}
\frac{\delta \! \stackrel{{\ldots }}{\phi ^{a}}\! \! (s)}{\delta \phi ^{b}(s')}=-2(\bff \bg )\delta (s-s')+\ldots 
\end{equation}

Finally, it remains to relate \( \delta \! \stackrel{{\ldots }}{\phi ^{a}}\! /\delta \phi ^{b} \)
and \( \delta g^{a}/\delta \phi ^{b} \). This relation is not immediate, because
the normal variation does not preserve the static gauge (\ref{c1}). The perturbed
surface
\begin{equation}
\label{d8}
\left\{ \begin{array}{l}
\tilde{x}_{0}(t,s)=t+\phi ^{a}(t,s)n^{a}_{0}(t,s)\: ,\\
\tilde{\bx }(t,s)=\bx (t,s)+\phi ^{a}(t,s)\bn ^{a}(t,s)\: 
\end{array}\right. 
\end{equation}
 has to be reparametrized by introducing the new coordinate \( \tilde{t}=\tilde{x}_{0} \),
so that 
\begin{equation}
\label{d9}
t=\tilde{t}-\phi ^{a}(\tilde{t},s)n^{a}_{0}(\tilde{t},s)+O(\phi ^{2})\: .
\end{equation}
 In the new coordinates the perturbed surface takes the static gauge form (we
rename \( \tilde{t}\to t \) )
\begin{equation}
\label{3.12}
\left\{ \begin{array}{l}
\tilde{x}_{0}(t,s)=t\: ,\\
\tilde{\bx }(t,s)=\bx (t,s)+\phi ^{a}(t,s)\bn ^{a}(t,s)-(\bff t+\bg t^{2})\phi ^{a}(t,s)n^{a}_{0}(t,s)\: .
\end{array}\right. 
\end{equation}

To use (\ref{3.12}), we must explicitly find \( n^{a}(t,s) \) satisfying (\ref{c16}).
The result is non-unique if \( D>2 \). Our choice was to first transform \( n^{a}(0,s)\Rightarrow \tilde{n}^{a}(t,s) \)
for each \( a \) by adding a linear combination of \( \dot{x}(t,s) \), \( x'(t,s) \)
so that \( \tilde{n}^{a} \) are orthogonal to the surface, and then to orthonormalize
the family \( \{\tilde{n}^{a}\} \). The result is: 
\begin{equation}
\label{d10}
\left\{ \begin{array}{l}
n^{a}_{0}(t,s)=-f_{a}t-g_{a}t^{2}+O(t^{3})\: ,\\
\bn ^{a}(t,s)=\bn ^{a}+\frac{1}{2}\bn ^{b}(f_{a}f_{b}t^{2}+(f_{a}g_{b}+f_{b}g_{a})t^{3})\\
\qquad \qquad -\bff f_{a}t^{2}-(\bg f_{a}+\bff g_{a})t^{3}+\bc 'O(t^{2})+O(t^{4})
\end{array}\right. 
\end{equation}
(all functions in the RHS are taken at \( s \)). Substituting this into (\ref{3.12})
gives
\begin{equation}
\label{3.13}
\delta g^{a}=\frac{1}{2}\stackrel{{\ldots }}{\phi ^{a}}+\frac{3}{2}(f_{a}g_{b}+f_{b}g_{a})\phi ^{b}\: .
\end{equation}
 Now it follows from (\ref{c38}) and (\ref{3.13}) that 
\begin{equation}
\label{d11}
\frac{\delta g^{a}(s)}{\delta \phi ^{a}(s')}=(4-D)\delta (s-s')+\ldots \: ,
\end{equation}
and hence \( \gamma =D-4 \) by (\ref{c15}).

\end{document}